\documentclass{webofc}
\usepackage[varg]{txfonts}   % Web of Conferences font
\begin{document}
\title{Bottom-hadron production in high-energy $pp$ and heavy-ion collisions}

\author{\firstname{Min} \lastname{He} \thanks{\email{minhephys@gmail.com}}
}

\institute{Department of Applied Physics, Nanjing University of Science and Technology, Nanjing 210094, China}

\abstract{%
The hadro-chemistry of bottom quarks produced in hadronic collisions encodes valuable information on the mechanism of color-neutralization in these reactions. We first compute the chemistry of bottom-hadrons in high-energy $pp$ collisions employing statistical hadronization with a largely augmented set of states beyond the currently measured spectrum. This enables a comprehensive prediction of fragmentation fractions of weakly decaying bottom hadrons for the first time and a satisfactory explanation of the existing measurements in $pp$ collisions at the LHC. Utilizing the bottom hadro-chemistry thus obtained as the baseline, we then perform transport simulations of bottom quarks in the hot QCD matter created in PbPb collisions at the LHC energy and calculate the pertinent bottom-hadron observables. The transverse momentum ($p_T$) dependent modifications of the bottom baryon-to-meson ratio ($\Lambda_b^0/B^-$) relative to their $pp$ counterparts are highlighted as a result of bottom quark diffusion and hadronization in the Quark-Gluon Plasma (QGP). We finally summarize the heavy quark (charm vs bottom) diffusion coefficients as extracted from transport simulations and compare them to result from recent full lattice QCD computations.
}
\maketitle
\section{Introduction}
\label{intro}
The production of bottom ($b$) quarks is believed to be well separated from the subsequent hadronization, as facilitated by its large mass. While the former can be reliably computed using perturbative QCD techniques, the hadronization as an intrinsically nonperturbative process relies on phenomenological modelling. The fragmentation fractions of $b$ quarks into weakly decaying $b$-hadrons (including strong and electromagnetic feeddown contributions from excited states), $f_u$, $f_d$, $f_s$ and $f_{\rm baryon}$, characterizing the probabilities of a $b$ quark hadronizing into a $B^-$, $\bar{B}^0$ and $\bar{B}_s^0$ meson and a $b$ baryon, respectively, thus provide a critical test of the hadronization mechanisms. These fractions have been measured in $p\bar{p}$ collisions at the Tevatron and found to be different from the values measured in $e^+e^-$ collisions~\cite{HFLAV:2019otj}, in that the $f_{\rm baryon}$ is significantly larger in hadronic collisions, implying that the $b$ quark hadronization is non-universal across different collision systems. The corresponding fractions in the charm sector have been measured by ALICE collaboration in $pp$ collisions at the LHC energies and similar enhancement of the baryon fraction relative to the $e^+e^-$ case was identified~\cite{{ALICE:2021dhb}}, confirming that the heavy quark hadronization may depend on the collision environment.

\section{Bottom hadro-chemistry in $pp$ collisions}
\label{sec-1}

\begin{figure}[h]
\centering
\includegraphics[width=8cm,clip]{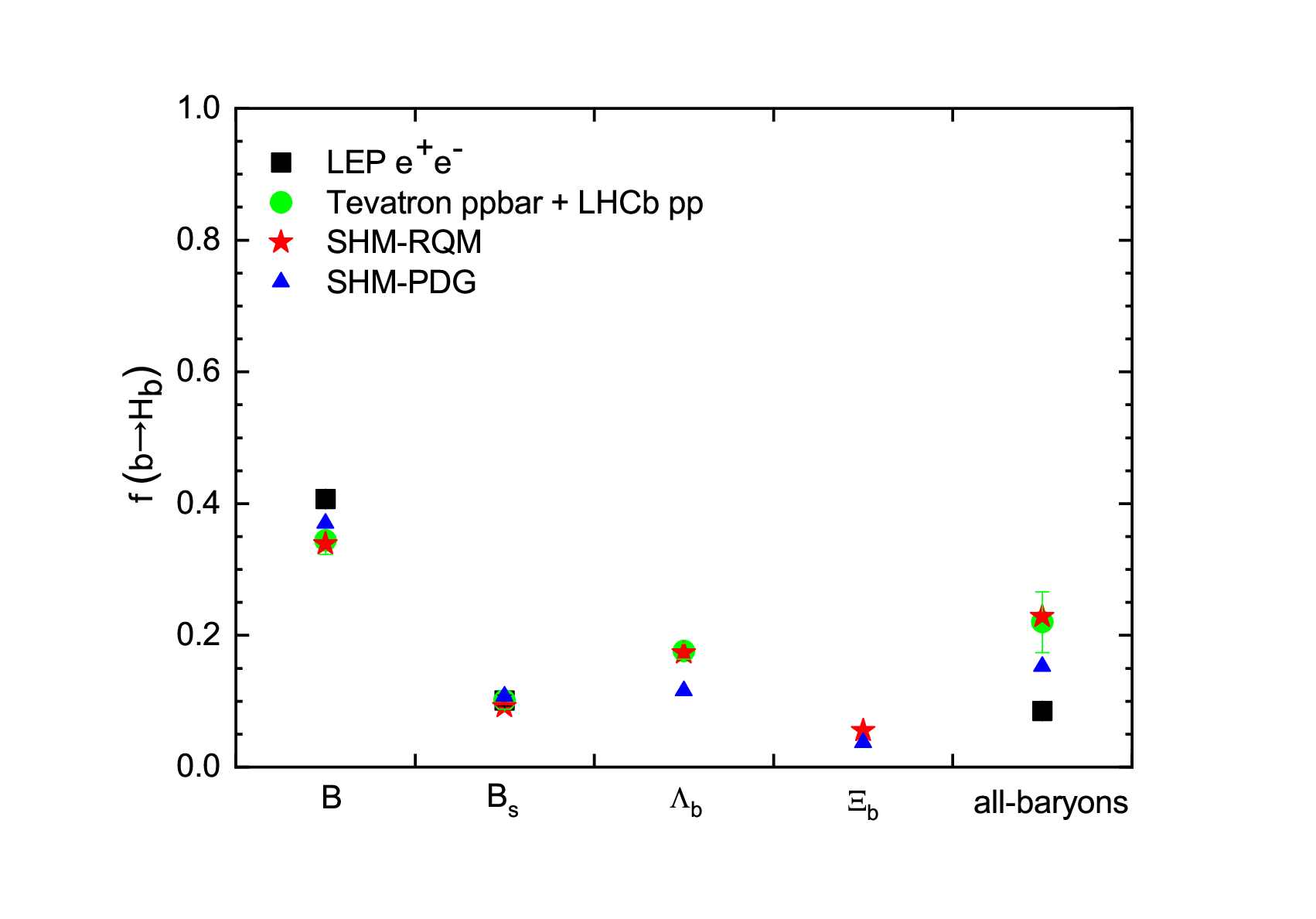}
\caption{Fragmentation fractions of ground state $b$ hadrons calculated from statistical hadronization model~\cite{He:2022tod}, in comparison with experimental measurements in $e^+e^-$ and $p\bar{p}$ or $pp$ collisions~\cite{HFLAV:2019otj}.}
\label{fig-fb-Hb}
\end{figure}

Assuming {\it relative} chemical equilibrium between different $b$-hadron production yields, we have evaluated the comprehensive set of fragmentation fractions of all weakly decaying $b$ hadrons in a generalized statistical hadronization model (SHM)~\cite{He:2022tod}. The thermal density of a given $b$ hadron of mass $m_i$ and spin-degeneracy $d_i$ and containing $N_s^i$ strange or antistrange quarks is computed at the hadronization temperature $T_H\sim170$\,MeV in the grand-canonical statistical ensemble
\begin{align}
n_i^{\rm primary}=\frac{d_i}{2\pi^2}\gamma_s^{N_s^i}m_i^2T_{H}K_2(\frac{m_i}{T_H}) \ ,
\label{primary_density}
\end{align}
with $\gamma_s\sim0.6$ being the strangeness suppression factor in elementary collisions. With the branching ratios (BR's) estimated from the $^3P_0$ quark model~\cite{He:2022tod} for the excited $b$ hadrons decaying into the ground state particles, the total densities of the weakly decaying $b$ hadrons are obtained via
\begin{equation}
n_{\alpha}=n_{\alpha}^{\rm primary} + \sum_in_i^{\rm primary}\cdot Br(i\rightarrow\alpha).
\label{total_density}
\end{equation}
Assuming that fragmentation weights of $b$ quarks into a given $b$-hadron is governed by its thermal density, Eq.~(\ref{total_density}) can be converted into the fragmentation fractions of the ground state $b$ hadrons, under the constraint of $f_u+f_d+f_s+f_{\Lambda_b^0}+f_{\Xi_b^{0,-}}+f_{\Omega_b^-}=1$. Fig.~\ref{fig-fb-Hb} summarizes the results for the fragmentation fractions of ground state $b$ hadrons from the statistical hadronization model with two scenarios for the input of $b$-hadron mass spectrum: the current particle data group (PDG) listings~\cite{ParticleDataGroup:2020ssz} and the relativistic quark model (RQM) predictions~\cite{Ebert:2009ua,Ebert:2011kk}. The fragmentation fractions as measured in hadronic ($pp$ and $p\bar{p}$) collisions are well reproduced by the SHM-RQM calculation, which features a significant enhancement of the baryonic fractions as a result of the feeddowns of the many yet unobserved excited states.

By further combining the transverse momentum ($p_T$) distributions of $b$ quarks from perturbative QCD (FONLL) calculations~\cite{Frixione:2007nw,Cacciari:2012ny} with thermal weights, we've conducted a fragmentation simulation
%using the fragmentation function
%\begin{equation}
%D_{b\rightarrow H_i}(z)\propto z^{\alpha}(1-z).
%\label{bFF}
%\end{equation}
which, complemented with the decay simulation, allows us to make predictions for the $p_T$ dependent ground-state baryon-to-meson ratio, $\Lambda_b^0/B^-$. As shown by the red curves in the left panel of Fig.~\ref{fig_pT-dependent-ratio-RAA}, the substantial gap between the LHCb data~\cite{LHCb:2019fns} and the SHM-PDG result is largely overcome by the feeddown of the large set of ``missing" baryons included in the SHM-RQM calculations.

\section{Bottom baryon-to-meson ratios in Pb-Pb collisions}

The hadro-chemistry computed above in $pp$ collisions serves as a controlled reference for studying its modifications in heavy-ion collisions.
To this end, we employ a strongly coupled transport approach previously developed for the charm sector~\cite{He:2019vgs} and calculate
the hadro-chemistry and nuclear modification factors of $b$ hadrons in $\sqrt{s_{NN}}$=5.02\,TeV PbPb collisions. This transport model features nonperturabtive interactions of heavy flavor with the medium, in terms of heavy quark diffusion in the QGP~\cite{Riek:2010fk}, resonance recombination as the dominant hadroniation mechanism at low to intermediate $p_T$~\cite{He:2022tod}, and heavy-hadron diffusion in the hadronic phase~\cite{He:2011yi}, therefore conceptually fully consistent with the notion of strongly-coupled QGP.

\begin{figure}[!t]
\centering
\hspace{-0.95cm}
\begin{minipage}{0.4\textwidth}
\includegraphics[width=1.2\textwidth]{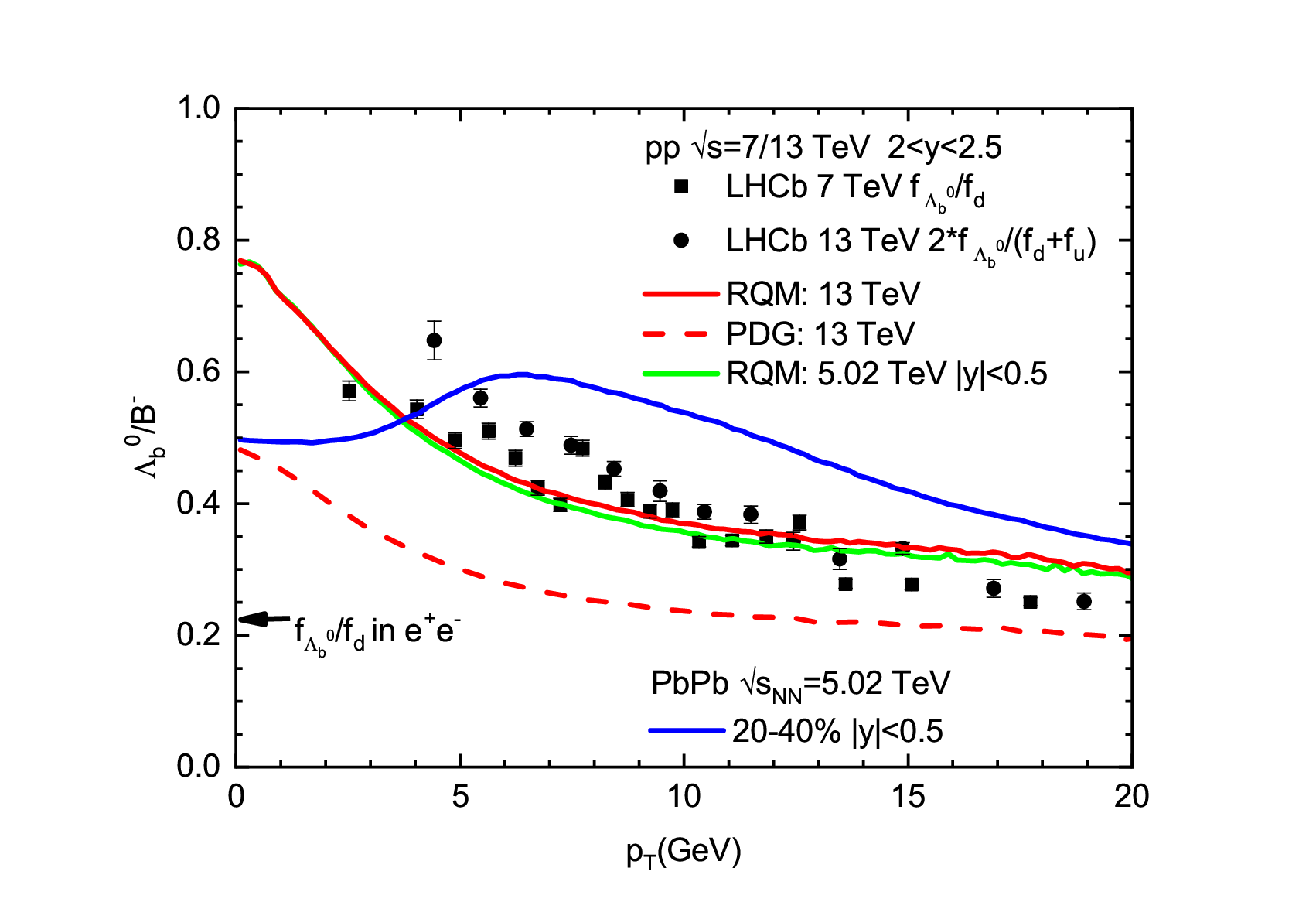}
\end{minipage}
\hspace{-0.5cm}
\begin{minipage}{0.4\textwidth}
\includegraphics[width=1.2\textwidth]{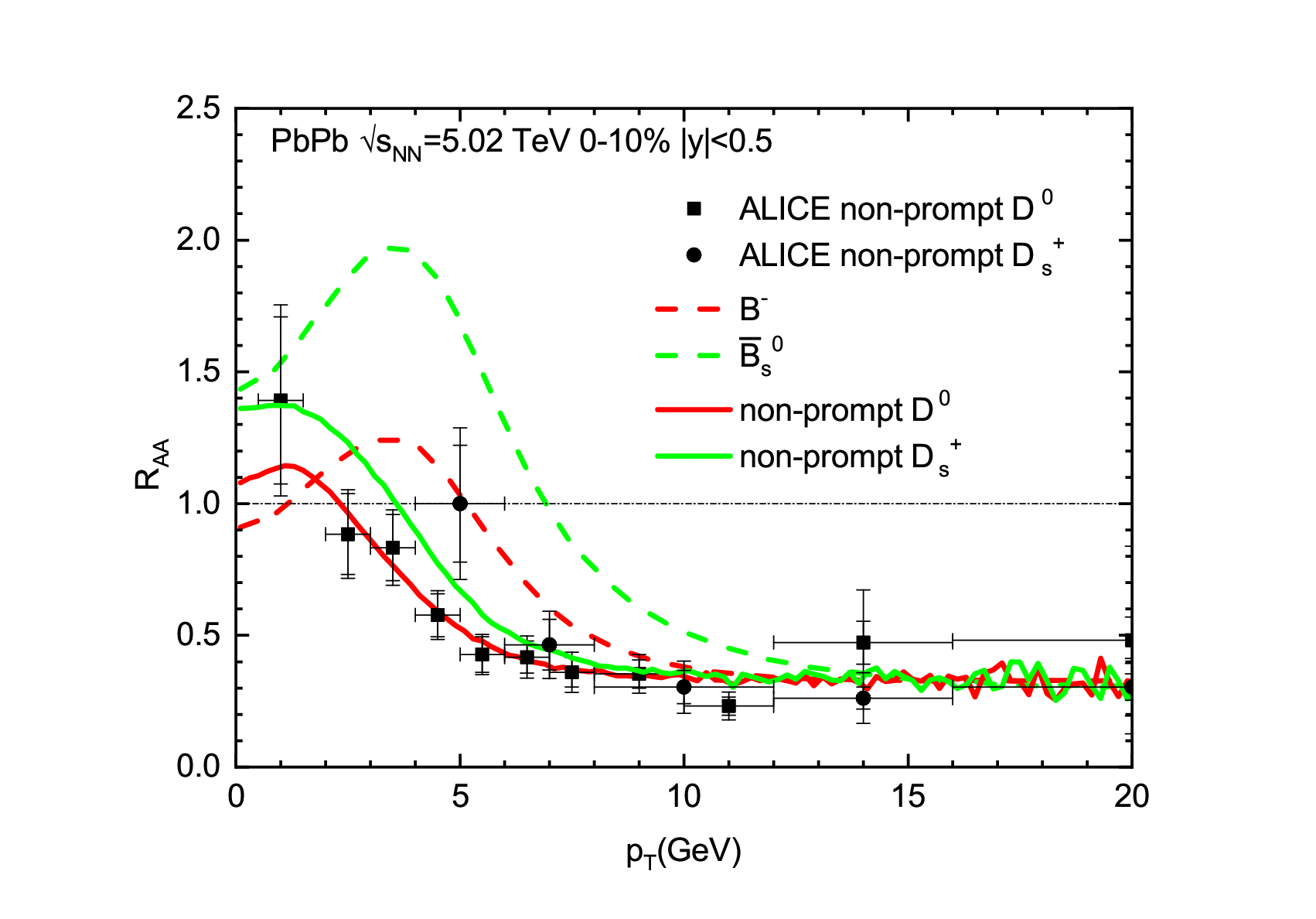}
\end{minipage}
\caption{Left: $\Lambda_b^0/B^-$ computed from SHM~\cite{He:2022tod} in comparison with LHCb measurements in $pp$ collisions~\cite{LHCb:2019fns}, and the same ratio computed from transport approach~\cite{He:2022tod} in 20-40\% $\sqrt{s_{NN}}$=5.02\,TeV PbPb collisions. Right: the nuclear modification factors~\cite{He:2022tod} for  non-prompt $D^0$ and $D_s^+$ (0-10\% PbPb collisions) in comparison with ALICE data~\cite{ALICE:2022tji,ALICE:2022xrg}.}
\label{fig_pT-dependent-ratio-RAA}
\end{figure}

We found that when amplifying the $b$-quark thermal relaxation rate as computed from $T$-matrix approach~\cite{Riek:2010fk} by a $K=1.6$ factor, which is to minic the missing contributions from radiative energy loss, the nuclear modification factors $R_{\rm AA}$ of both nonprompt $D$ and $D_s$ mesons, which are the weak decay product of $b$ hadrons including $b$ baryons, can be well reproduced, as shown in the right panel of Fig.~\ref{fig_pT-dependent-ratio-RAA}. Furthermore, the full set of $R_{\rm AA}$'s of ground-state $b$ hadrons develops an expected hierarchy of flow effects and suppression as driven by their quark content~\cite{He:2022tod}. Accordingly, the $\Lambda_b^0/B^-$ ratio exhibits a significant enhancement at intermediate-$p_T$ (blue curve in the left panel of Fig.~\ref{fig_pT-dependent-ratio-RAA}) relative to the $pp$ baseline due to a stronger flow effect on generally heavier baryons, which is fully captured by the resonance recombination incorporating space-momentum correlations~\cite{He:2022tod}. Notable is the peak of $\Lambda_b^0/B^-$  at a higher $p_T\sim$6\,GeV and extending to significantly larger $p_T\sim$15\,GeV than the corresponding ratio in the charm sector~\cite{He:2019vgs} because of the large $b$-quark mass.

\begin{figure}[h]
\centering
\includegraphics[width=8cm,clip]{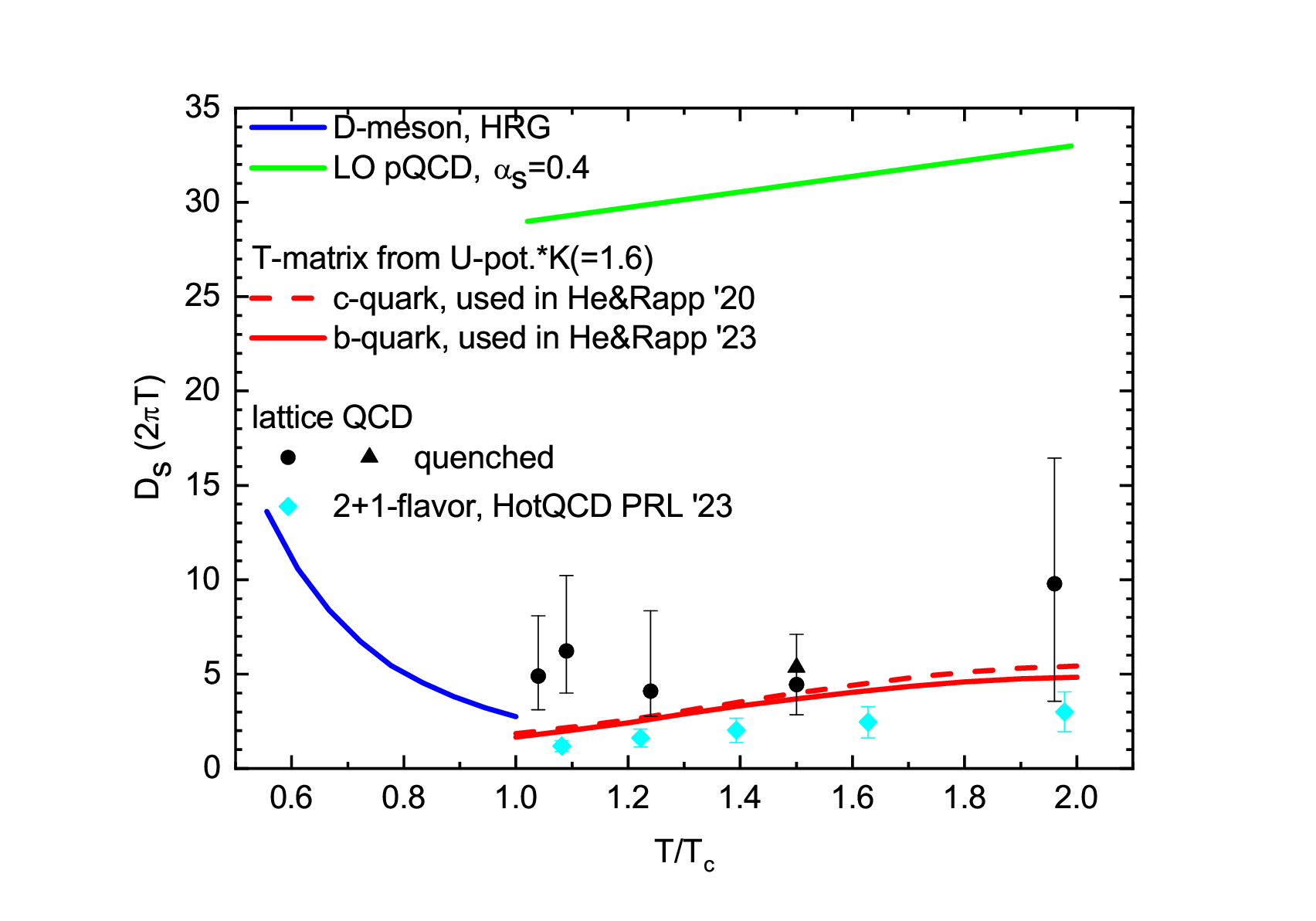}
\caption{Charm~\cite{He:2019vgs} and bottom~\cite{He:2022tod} quark spatial diffusion coefficient in unit of thermal wave length as extracted from the nonperturbative transport model calculations of charm- and bottom-hadron observables in comparison with experimental data in PbPb collisions. The recent full lattice QCD result~\cite{Altenkort:2023oms} is also plotted for comparison. The $D$-meson diffusion coefficient (blue line) is taken from~\cite{He:2011yi}.}
\label{fig-Ds2piT}
\end{figure}

In Fig.~\ref{fig-Ds2piT}, the charm and bottom quark spatial diffusion coefficients in unit of thermal wave length as used in~\cite{He:2019vgs} and ~\cite{He:2022tod}, respectively, representing decent description of both charm and bottom hadro-chemistry, collective flow and suppression observables, are summarized. The extracted charm and bottom quark diffusion coefficients are almost identical and come to a small value of $D_s(2\pi T)\sim$2-4 at temperatures (1-1.5$T_c$) near the phase boundary, implying a strong coupling of heavy quarks with the medium. It is remarkable to find that the heavy quark diffusion coefficient extracted from nonperturabtive transport model calculations very comparable to recent full lattice QCD result~\cite{Altenkort:2023oms}.

\section{Conclusions}
The bottom hadrochemistry in $pp$ collisions has been addressed in the generalized statistical hadronization model, corroborating that a vast spectrum of additional $b$-baryon excited states awaits discovery. Using this as a baseline, the $b$-hadron collective flow patten in PbPb collisions has been computed in a nonperturbative transport model and a small heavy quark diffusion coefficient has been extracted.


\begin{thebibliography}{0}

\end{thebibliography}


\begin{thebibliography}{}


\bibitem{HFLAV:2019otj}
Y.~S.~Amhis \textit{et al.} [HFLAV],
%``Averages of b-hadron, c-hadron, and $\tau $-lepton properties as of 2018,''
Eur. Phys. J. C \textbf{81}, no.3, 226 (2021).
%doi:10.1140/epjc/s10052-020-8156-7
%[arXiv:1909.12524 [hep-ex]].
%854 citations counted in INSPIRE as of 31 Dec 2023

%\cite{ALICE:2021dhb}
\bibitem{ALICE:2021dhb}
S.~Acharya \textit{et al.} [ALICE],
%``Charm-quark fragmentation fractions and production cross section at midrapidity in pp collisions at the LHC,''
Phys. Rev. D \textbf{105}, no.1, L011103 (2022).
%doi:10.1103/PhysRevD.105.L011103
%[arXiv:2105.06335 [nucl-ex]].
%96 citations counted in INSPIRE as of 31 Dec 2023


%\cite{He:2022tod}
\bibitem{He:2022tod}
M.~He and R.~Rapp,
%``Bottom Hadrochemistry in High-Energy Hadronic Collisions,''
Phys. Rev. Lett. \textbf{131}, no.1, 1 (2023).
%doi:10.1103/PhysRevLett.131.012301
%[arXiv:2209.13419 [hep-ph]].
%5 citations counted in INSPIRE as of 31 Dec 2023


\bibitem{ParticleDataGroup:2020ssz}
P.~A.~Zyla \textit{et al.} [Particle Data Group],
%``Review of Particle Physics,''
PTEP \textbf{2020}, no.8, 083C01 (2020).
%doi:10.1093/ptep/ptaa104

\bibitem{Ebert:2009ua}
D.~Ebert, R.~N.~Faustov and V.~O.~Galkin,
%``Heavy-light meson spectroscopy and Regge trajectories in the relativistic quark model,''
Eur. Phys. J. C \textbf{66}, 197-206 (2010).
%doi:10.1140/epjc/s10052-010-1233-6
%[arXiv:0910.5612 [hep-ph]].


\bibitem{Ebert:2011kk}
D.~Ebert, R.~N.~Faustov and V.~O.~Galkin,
%``Spectroscopy and Regge trajectories of heavy baryons in the relativistic quark-diquark picture,''
Phys. Rev. D \textbf{84}, 014025 (2011).
%doi:10.1103/PhysRevD.84.014025
%[arXiv:1105.0583 [hep-ph]].

\bibitem{Frixione:2007nw}
S.~Frixione, P.~Nason and G.~Ridolfi,
%``A Positive-weight next-to-leading-order Monte Carlo for heavy flavour hadroproduction,''
JHEP \textbf{09}, 126 (2007).
%doi:10.1088/1126-6708/2007/09/126
%[arXiv:0707.3088 [hep-ph]].

\bibitem{Cacciari:2012ny}
M.~Cacciari {\it et al.,}
%``Theoretical predictions for charm and bottom production at the LHC,''
JHEP \textbf{10}, 137 (2012).
%doi:10.1007/JHEP10(2012)137
%[arXiv:1205.6344 [hep-ph]].

%\cite{LHCb:2019fns}
\bibitem{LHCb:2019fns}
R.~Aaij \textit{et al.} [LHCb],
%``Measurement of $b$ hadron fractions in 13 TeV $pp$ collisions,''
Phys. Rev. D \textbf{100}, no.3, 031102 (2019).
%doi:10.1103/PhysRevD.100.031102
%[arXiv:1902.06794 [hep-ex]].
%143 citations counted in INSPIRE as of 01 Jan 2024


\bibitem{ALICE:2022tji}
S.~Acharya \textit{et al.} [ALICE],
%``Measurement of beauty production via non-prompt D$^{0}$ mesons in Pb-Pb collisions at $ \sqrt{{\textrm{s}}_{\textrm{NN}}} $= 5.02 TeV,''
JHEP \textbf{12}, 126 (2022).
%doi:10.1007/JHEP12(2022)126
%[arXiv:2202.00815 [nucl-ex]].

%\cite{ALICE:2022xrg}
\bibitem{ALICE:2022xrg}
S.~Acharya \textit{et al.} [ALICE],
%``Measurement of beauty-strange meson production in Pb\textendash{}Pb collisions at sNN=5.02TeV via non-prompt Ds+ mesons,''
Phys. Lett. B \textbf{846}, 137561 (2023).
%doi:10.1016/j.physletb.2022.137561
%[arXiv:2204.10386 [nucl-ex]].
%12 citations counted in INSPIRE as of 01 Jan 2024

%\cite{He:2019vgs}
\bibitem{He:2019vgs}
M.~He and R.~Rapp,
%``Hadronization and Charm-Hadron Ratios in Heavy-Ion Collisions,''
Phys. Rev. Lett. \textbf{124}, no.4, 042301 (2020).
%doi:10.1103/PhysRevLett.124.042301
%[arXiv:1905.09216 [nucl-th]].


\bibitem{Riek:2010fk}
F.~Riek and R.~Rapp,
%``Quarkonia and Heavy-Quark Relaxation Times in the Quark-Gluon Plasma,''
Phys. Rev. C \textbf{82}, 035201 (2010).
%doi:10.1103/PhysRevC.82.035201
%[arXiv:1005.0769 [hep-ph]].


\bibitem{He:2011yi}
M.~He, R.~J.~Fries and R.~Rapp,
%``Thermal Relaxation of Charm in Hadronic Matter,''
Phys. Lett. B \textbf{701}, 445-450 (2011).
%doi:10.1016/j.physletb.2011.06.019
%[arXiv:1103.6279 [nucl-th]].

%\cite{Altenkort:2023oms}
\bibitem{Altenkort:2023oms}
L.~Altenkort \textit{et al.} [HotQCD],
%``Heavy Quark Diffusion from 2+1 Flavor Lattice QCD with 320~MeV Pion Mass,''
Phys. Rev. Lett. \textbf{130}, no.23, 231902 (2023).
%doi:10.1103/PhysRevLett.130.231902
%[arXiv:2302.08501 [hep-lat]].
%25 citations counted in INSPIRE as of 01 Jan 2024

\end{thebibliography}
\end{document}